\documentclass[11pt]{article}

\usepackage{amssymb,amsmath}                                           % Standard math packages
\usepackage{bbm}                                                       % Fancy blackborld bold
\usepackage{tensor}                                                    % For fancy indices
\usepackage[usenames,dvipsnames]{color}                                % For colors (for hyperlinks)
\usepackage{setspace}                                                  % For spacing,  must come BEFORE `hyperref' package.
\usepackage{hyperref}                                                  % For hyperlinks
\usepackage[left= 1in,right=1in,top=1in,bottom=1in]{geometry}          % For margin settings
\usepackage{comment}                                                   % For comment environment
\usepackage{authblk}                                                   % For author affiliations, etc...
\usepackage{graphicx}                                                  % For ``includegraphics''
\usepackage{caption}                                                    % For fancy captions and subcaptions.
\usepackage{subcaption}

\hypersetup{                 
    colorlinks=true,         % false: boxed links; true: colored links, false is default
    linkcolor=blue,          % color of internal links, red is default
    citecolor=red,           % color of links to bibliography, 'green' is default
    urlcolor=Violet          % color of external links, cyan is default
}

\let\oldmarginpar\marginpar
\renewcommand\marginpar[1]{\oldmarginpar{\color{red}\raggedright\scriptsize #1}}
\newcommand{\pb}[2]{\ensuremath{\lf\{#1,#2 \rt\}}}
\newcommand{\mean}[1]{\ensuremath{\lf\langle #1 \rt\rangle }}
\newcommand{\diby}[2]{\ensuremath{\frac{\partial #1}{\partial #2}}}

\newcommand{\sfrac}{\textstyle \frac}

\def\lf {\ensuremath{\left}}
\def\rt {\ensuremath{\right}}

\def\ham {\ensuremath{\mathcal H}}
\def\gau {\ensuremath{\mathcal P}}

\def\de {\ensuremath{ {\rm d} }}
\def\Epsilon {\ensuremath{\mathcal E}}
\def\gone {\ensuremath{\text{Type~1}}}
\def\gtwo {\ensuremath{\text{Type~2}}}

\title{Schr\"odinger Evolution for the Universe: Reparametrization}

\author[1]{{\bf Sean Gryb}\thanks{email: \href{mailto:s.gryb@hef.ru.nl}{s.gryb@hef.ru.nl}}}\author[2]{\bf Karim Th\'ebault\thanks{email: \href{mailto:karim.thebault@gmail.com}{karim.thebault@gmail.com}}}
\affil[1]{ Radboud University Nijmegen, Institute for Mathematics, Astrophysics and Particle Physics, The~Netherlands}
\affil[2]{ Munich Center for Mathematical Philosophy, Ludwig Maximilians Universit\"{a}t, Ludwigstrasse 31, D-80539, Munich, Germany}

\date{\today}

\begin{document}

\maketitle

\begin{abstract}
Starting from a generalized Hamilton--Jacobi formalism, we develop a new framework for constructing observables and their evolution in theories invariant under global time reparametrizations. Our proposal relaxes the usual Dirac prescription for the observables of a totally constrained system (`perennials') and allows one to recover the influential partial and complete observables approach in a particular limit. Difficulties such as the non-unitary evolution of the complete observables in terms of certain partial observables are explained as a breakdown of this limit. Identification of our observables (`mutables') relies upon a physical distinction between gauge symmetries that exist at the level of histories and states (`Type 1'), and those that exist at the level of histories and not states (`Type 2'). This distinction resolves a tension in the literature concerning the physical interpretation of the partial observables and allows for a richer class of observables in the quantum theory. There is the potential for the application of our proposal to the quantization of gravity when understood in terms of the Shape Dynamics formalism.

\end{abstract}

\clearpage
\tableofcontents

\section{Introduction}

\subsection{Observables and Change in Totally Constrained Systems}

Identification of the physical quantities that can be measured -- the `observables' -- within a constrained Hamiltonian theory generally follows a prescription, due to Dirac \cite{dirac:lectures}, whereby the theory is restricted to functions on the physical phase space that  (weakly) commute with the first class constraints. The various  quantization techniques applicable to these theories are then required to preserve this notion of observable in the quantum context \cite{Henneaux:1992a}. However, as is well-known, the use of \textit{Dirac observables} is problematic in totally constrained systems such as general relativity within which the Hamiltonian is itself a constraint. In such theories, the Dirac observables are unchanging `perennials' and, thus, there is an acute problem in recovering a description of physical change. Several promising lines of argument towards the construction and interpretation of perennials exist in the literature. In particular, the partial and complete observables approach \cite{Rovelli:2002,Rovelli:2004,Rovelli:2007,Dittrich:2006, Dittrich:2007, Thiemann:2007} provides an algorithm for the construction of Dirac observables in totally constrained systems.   

Here we take a different approach to the problem of observables and change in totally constrained systems, but one which contains the Dirac observables as a special case. On our view, one must distinguish between two notions of gauge invariance often taken as equivalent in the literature. The first notion was emphasised in Dirac's original work \cite{dirac:lectures} and arises when there exists an equivalence class of \emph{states} in the state space of a theory that are identified as \emph{physically indistinguishable}, leading to an under-determination problem in the evolution equations of the theory. A second notion of gauge invariance  arises when there is an equivalence class of \emph{histories} in the space of allowable histories of a theory.

A  symmetry at the level of states in the state space is not a priori dependent upon the invariance of the action. Rather, such symmetries result directly from redundancy within the representation of the freely specifiable initial data of the system. As an example, take the case of the `hidden' Weyl invariance of General Relativity that becomes manifest when expressed in terms Shape Dynamics \cite{gryb:shape_dyn}. In this case, a parametrization of the gauge-invariant degrees of freedom of the theory can be found in a special choice of foliation where the physical degrees of freedom can be expressed solely in terms of Weyl invariant quantities, even though this Weyl invariance is not an explicit symmetry of the action of general relativity.

 A symmetry at the level of histories, on the other hand, is defined directly in terms of an invariance of the action. Any state symmetry implies the existence of a history symmetry as a special case, since a history is just a sequence of states. However, the converse is not true. Even in well-known cases of gauge theories, such as Maxwell theory or Yang--Mills theory, care must be taken to ensure that a state symmetry is equivalent to a history symmetry when the gauge transformations are time-dependent \cite{Pons:1997,PonsSalis:2005,Pons:2010,pitts:2013,pitts:2014}.

The distinction between gauge symmetries occurring at the level of `histories \emph{and} states' and gauge symmetries occurring at the level of `histories \emph{but not} states' will be of fundamental importance to the arguments of this paper. We will therefore make the following definitions: a `$\gone$' symmetry exists within a theory when there is a state symmetry with a corresponding history symmetry; a `$\gtwo$' symmetry exists when there is history symmetry with \emph{no} corresponding state symmetry.\footnote{For an exploration of the basis behind these two notions in terms of the distinction between `manifest free' and `manifest fixed' variations of the action see \cite{gryb:2014}.} Formally, it is possible to distinguish between $\gone$ and $\gtwo$ by looking at the infinitesimal action \textit{on configuration space} induced by the \textit{phase space} constraint associated with the symmetry in question. For $\gone$ symmetries, the constraint will induce a well-defined action on configuration space. This is because $\gone$  symmetries can be understood as transformations of instantaneous states in isolation, i.e., they need not make reference to the particular history to which a state belongs. $\gtwo$ symmetries, in contrast, require information about the history a particular state belongs to in order to be full specified. This means that the there is no phase space constraint function that induces a well-defined action on configuration space corresponding to the $\gtwo$ symmetry transformation. The prototypical example of a $\gone$ symmetry is the $U(1)$ invariance of Maxwell theory while, for a $\gtwo$ symmetry, the prototypical example is reparametrization invariance.

One benefit of the distinction between $\gone$ and $\gtwo$ relates to the complete and partial observable program mentioned above. This approach involves us  identifying both a set of phase space variables -- the `partial observables' -- and a set of correlations between these observables -- the `complete observables'. Whilst it is generally agreed that the complete observables are measurable quantities, the status of the partial observables is controversial. Whereas Rovelli \cite{Rovelli:2002,Rovelli:2004,Rovelli:2007,Rovelli:2014} identifies these functions with physical quantities that can be measured, Thiemann \cite{Thiemann:2007} argues that partial observables should be understood as gauge-dependent quantities that are non-measurable. As shall be detailed below, there are good reasons to endorse the Thiemann perspective for partial observables connected to $\gone$ symmetries and the Rovelli perspective for partial observables connected to $\gtwo$ symmetries.

A more significant insight gained from the $\gone$ and $\gtwo$ distinction relates to the quantization of reparametrization invariant theories with a global notion of time. The Dirac quantization algorithm implements a kinematical prescription to remove the dependence of both the wavefunction and the observables on the parameters that label the `gauge orbits' connecting physically indistinguishable states. Dirac's prescription is, thus, explicitly motivated by the identification of the instantaneous states along the gauge orbits. Although such an identification is justified in the case of a $\gone$ symmetry, it is certainly \emph{not} justified in the $\gtwo$ case. Thus, on our view, the Dirac quantization algorithm is   only justified for the $\gone$ case. One then requires a new definition of observables and a new approach to quantization for gauge theories with $\gtwo$ symmetries. This is precisely what the \emph{relational quantization} program, defended here and elsewhere \cite{gryb:2011,gryb:2014}, aims to provide.\footnote{`Relational quantization', in our terminology, implies an approach towards the quantization of totally constrained systems within which fundamental time ordering structure is retained. We view any approach that jettisons this structure as `timeless' rather than properly speaking `relational'. In this respect we are proposing a brand of `quantum relationalism' that is conceptually distinct from others approaches in the literature. See \cite{Page:1983,Rovelli:2004,Bojowald:2011,Anderson:2012} for more on other notions of `relationalism' explored in the quantum context and \cite{Gryb:2015} for more on the philosophical background to our approach.}     

Below we will describe a new implementation of relational quantization and show how it correctly implements Dirac's requirements when, and only when, they are appropriate. Whilst, in previous work, relational quantization was presented in terms of path integral \cite{gryb:2011} or constraint quantization techniques \cite{gryb:2014}, here we will emphasise the connection to a generalised Hamilton--Jacobi formalism. One particular feature of this new formalism is that it allows us to make a direct connection between the change of the classical characteristic functional with respect to an independent parameter, and the change of the quantum wavefunction with respect to the same parameter. When such a parameter is associated with a $\gtwo$ symmetry, we take both the classical and quantum changes to be physical. The change that is parameterized is treated as physical even though the independent parameter labelling the change does not correspond to a measurable quantity. Reparametrization invariance in the presence of a global time label is the most important example of a $\gtwo$ symmetry and, in that case, the relevant change parameter is time. Thus, relational quantization leads to a quantum reparametrization invariant formalism with genuine dynamical change. This is in stark contrast to quantum theories resulting from all other canonical quantization techniques. Within the quantum formalism of standard approaches, a Wheeler--DeWitt-type equation imposes the \textit{kinematical} restriction of invariance of the wavefunction with respect the time parameter. Within a relationally quantized theory, a Schr\"{o}dinger-type equation gives the \textit{dynamical} evolution of the wavefunction with respect the time parameter. 

Since change with regard to the time parameter is treated as physical, relational quantization allows for the possibility of such change within the observables of a totally constrained theory. Our approach implements a dynamical notion of observable that is closely related to that defended by Kucha\v{r} \cite{Kuchar:1991,Kuchar:1992} (see also \cite{barbour:2008}). Like Kucha\v{r}, we disagree with the `perennials' prescription whereby only the timeless Dirac observables of a totally constrained system have the capacity to represent physical quantities. And, like Kucha\v{r}, we endorse an enlarged set of observables that need not commute with all the constraints. However, our prescription is distinct from that of Kucha\v{r} since we are more restrictive as to the types of constraints that observables may be non-commuting with respect to.

Let us introduce three notions of observable relevant to totally constrained systems as follows: \textit{Perennials} are  functions that commute with all the (first class) constraints; \textit{Mutables} are functions that commute with the (first class) constraints associated with $\gone$ symmetries but not those associated with $\gtwo$ symmetries; \textit{Kucha\v{r} observables} commute with all the (first class) constraints apart from Hamiltonian constraints. When there is a single Hamiltonian constraint, this constraint is associated with a $\gtwo$ symmetry and the Kucha\v{r} observables and  mutables are identical. When there are multiple Hamiltonian constraints, in theories such as canonical general relativity, things become more complicated and the two notions come apart.

\subsection{The Problem of Refoliations}

The infinite family of Hamiltonian constraints in canonical general relativity are connected to \textit{local} time diffeomorphism invariance. Crucially, local time diffeomorphism invariance cannot be unambiguously classified as either a $\gone$ or $\gtwo$ symmetry. This is because the set of local time diffeomorphisms include two very different types of transformations that are not clearly distinguished in the formalism.  Canonical general relativity involves specification of geometrical information relating to both canonical data on sequences of spatial hypersurfaces and the embeddings of these hypersurfaces into spacetimes. Local time diffeomorphisms are a set of symmetry transformations that includes both transformations that preserve embeddings (reparametrizations) and transformations that do not (refoliations). The embedding preserving transformations can be unambiguously understood as $\gtwo$ symmetries: it makes sense to talk of reparametrizations as symmetries of histories but not states since they reparametrize phase space curves without changing their image. Refoliations, on the other hand, change the image of phase space curves (like a $\gone$ symmetry) but do so without admitting a unique state-by-state representation of their action. This is because, in addition to a history, one needs to specify the embedding of the phase space data into spacetime in order to define a refoliation.  Without this embedding information, it is impossible to compare two histories on phase space. Another way of viewing this issue is to consider refoliations as normal deformations of three dimensional hypersurfaces embedded within four geometries \cite{Teitelboim:1973}. In that context, it is clear that they will require a spacetime metric in order to be defined. This metric can be specified either in terms of spacetime geometric data or via integration of spatial data \emph{and} the necessary smearing functions that specify the embedding. Under either viewpoint, we see that, by definition, refoliations cannot be classified as either $\gone$ or $\gtwo$ symmetries. This suggests the identification of a new `Type~3' symmetry consisting of transformations that: i) are symmetries of the action; ii) do not preserve embeddings; and iii) have no unambiguous action on phase space. 

Since phase space curves map to curves on configuration space, there are obvious difficulties in the quantization of Type~3 symmetries. In particular it is difficult to see how one can consistently define a wavefunction in a configuration basis unless one also has a way to specify the appropriate embedding information quantum mechanically. We will not attempt to tackle this problem here. Rather, given that refoliation invariance is a symmetry of Type~3, we accept that relational quantization, as well as the associated arguments regarding observables and change, cannot be directly applied to gravity when expressed in the full refoliation invariant formalism of canonical general relativity. Fortunately, considerations regarding the interpretation of scale suggest an alternative formulation for gravity where instantaneous states can be used to construct complete solutions \emph{without} having to specify the embedding of the hypersurfaces into spacetime. Instead, this embedding can be constructed from the scale invariance alone. Because this new interpretation of gravity --- called \emph{Shape Dynamics} \cite{ barbour_el_al:physical_dof, Barbour:new_cspv, gryb:shape_dyn, Gomes:linking_paper} --- provides its own embedding, its symmetries can be unambiguously divided into $\gone$ and $\gtwo$, and relational quantization can be applied. In this paper, we will lay the necessary groundwork for considering the case of gravity by showing how relational quantization leads to an explicit characterisation of mutables in models with global reparametrization symmetry. In these cases, it is demonstrated that perennials constructed via the partial and complete observables scheme can be recovered as a subset of the mutables constructed within our approach. Thus, the formalism presented here allows us to: i) give an  explicit construction of what constitutes an observable in a relationally quantized framework, and ii) make direct contact between our previous proposal and `internal time' methods for understanding change within fully constrained Hamiltonian systems.

\section{Generalized Hamilton--Jacobi Formalism} \label{sec:general formalism}

We begin by presenting a generalized Hamilton--Jacobi formalism for constrained classical systems. This approach is closely related to those found in the literature \cite{Rovelli:2001,Rovelli:2004,Souriau:1997}.  In particular, for $\gone$ symmetries, our formalism reproduces the usual Dirac formalism; however, in the case of $\gtwo$ symmetries, there are several crucial differences. An explicit application to a particle model that features both globally reparametrization-invariant symmetries, which we take to be $\gtwo$, and $\gone$ symmetries will be provided in the following section.

Consider a totally constrained Hamiltonian theory on a phase space $\Gamma$, coordinatized in some chart by $(q,p)$, defined by the first class system of $r$ constraints $C_\alpha(q,p)\approx 0$, for $\alpha=1,...r$. The canonical action reads
\begin{equation}
	S = \int \lf[ p\cdot  \dot q - \lambda^\alpha C_\alpha(q,p) \rt]\de t\,.
\end{equation}
For the moment, we use an abstract index-free notion for phase space variables $(q,p)$ so that the dot product represents an abstract inner product on $\Gamma$. Thus, our considerations will generally apply to the infinite dimensional case. Integration is over the arbitrary parameter $t$ and over-dots represent $t$-derivatives. The symplectic term requires $\Gamma$ to be equipped with the symplectic potential $\theta = p \wedge \de q$ which defines a symplectic 2-form $\omega = \de \theta$ (where exterior derivatives and wedge products are defined on $\Gamma$). In addition, we will assume that $\Gamma$ is equipped with a canonical Poisson structure such that $\pb q p = \mathbbm 1$ in some chart and all other Poisson brackets are zero.

Our goal is to find a canonical transformation that parametrizes the flow of the constraints $C_\alpha$ locally on $\Gamma$ and allows us, in particular, to restrict this flow to the constraint surface defined by $C_\alpha \approx 0$. To do this, consider the  modified action:
\begin{equation}\label{eq:extended S}
	S_\text{e} = \int \lf[ p\cdot  \dot q + \dot\phi^\alpha \cdot \Epsilon_\alpha - \lambda^\alpha \lf( \Epsilon_\alpha - C_\alpha(q,p) \rt) \rt]\de t
\end{equation}
defined on the \emph{extended} phase space $\Gamma(q,p)\to \Gamma_\text{e}(q,p; \phi^\alpha, \Epsilon_\alpha)$ coordinatized in some finite patch by the additional variables $(\phi^\alpha, \Epsilon_\alpha)$. The symplectic potential on $\Gamma_\text{e}$ is now
\begin{equation}
	\theta_\text{e} = p \wedge \de q + \Epsilon_\alpha \wedge \de \phi^\alpha\,.
\end{equation}
We additionally assume the Poisson structure $\pb {\phi^\alpha}{\Epsilon_\beta} = \delta^\alpha_\beta$ so that $\phi^\alpha$ and $\Epsilon_\alpha$ are canonically conjugate. Eventually, we will use $\phi^\alpha$ to locally parametrize a canonical transformation along the orbits of $C_\alpha$. The momenta $\Epsilon_\alpha$ are easily seen to be constants of motion, since their equations of motion imply
\begin{equation}
	\dot \Epsilon_\alpha = \pb {\Epsilon_\alpha} {H} = 0\,.
\end{equation}
For the special initial condition $\Epsilon = 0$, we see that the extended theory defined by $S_\text{e}$ is classically equivalent to the original theory defined by $S$.

In order for the constraints $\Epsilon_\alpha - C_\alpha(q,p) \approx 0$ to form a first class surface on the extended phase space $\Gamma_\text{e}$, the functions $C(q,p)$ must be Abelian (i.e., $\pb{C_\alpha}{C_\beta} = 0$). For simplicity, we will assume that the $C(q,p)$'s have already been \emph{Abelianized}, which is always possible locally on $\Gamma$ due to Darboux's theorem \cite[p.175]{AbMa1978}.\footnote{Darboux's theorem states that, locally on $\Gamma$, one can always coordinatize the constraint surface $C_\alpha(p,q) = 0$ in terms of ``orthogonal'' (i.e., commuting) coordinates. When re-expressed in terms of these coordinates, the constraints are, by definition, Abelian.} Moreover, for the quantum theory, it is also possible to perform this Abelianzation globally (see for instance \cite{Gogilidze:1995ky}), although, in practice, this procedure can be exceedingly difficult to perform explicitly. Nevertheless, in principle, modulo a potentially messy Abelianization procedure, this assumption leads to no loss of generality. Fortunately, for the important case of gravity considered in our companion paper, relational quantization is applied to Shape Dynamics whose generator of evolution is Abelian with respect to all local constraints of the theory. Thus, the Abelian restriction is irrelevant in that case.

We are now in a position to define a canonical transformation that parametrizes the flow of the constraints. This can be achieved by requiring that the new coordinates $(P,Q; \Phi^\alpha, E_\alpha)$ have zero flow under the transformed $C_\alpha$, so that they are analogous to the `initial data' of standard Hamilton--Jacobi theory. For the old coordinates $(q,p)$, we additionally require that the momenta $\Epsilon_\alpha$ are constrained to be equal to the $C_\alpha(q,p)$. The new symplectic potential $\theta^\prime_\text{e} = P \wedge \de Q + E_\alpha \wedge \de \Phi^\alpha$ must differ from the old one by an exact form $\de S$. This ensures that the symplectic 2-form $\omega_\text{e} = \de \theta_\text{e} = \de \theta^\prime_\text{e}$ is invariant. Implementing the above requirements, we obtain
\begin{align}
	\de S(q,Q; \phi^\alpha, \Phi_\alpha) &= \theta_\text{e} - \theta^\prime_\text{e} \notag \\
	&= p\wedge \de q + \Epsilon_\alpha \wedge \de \phi^\alpha - P \wedge \de Q - E_\alpha \wedge \de \Phi^\alpha \notag \\
	&= p\wedge \de q + C_\alpha(p,q) \wedge \de \phi^\alpha - P \wedge \de Q\,,
\end{align}
where in the last line we set $\Epsilon_\alpha = C_\alpha(q,p) $ and $E_\alpha = 0 $. The reason for setting $E_\alpha = 0$ is because $E_\alpha$ represents the generator of flow on the transformed coordinates $(Q,P)$, which we want to vanish. This is completely analogous to requiring that the transformed Hamiltonian vanish in standard Hamilton--Jacobi theory.

\begin{figure}
	\begin{center}
		\includegraphics[width= .8\textwidth]{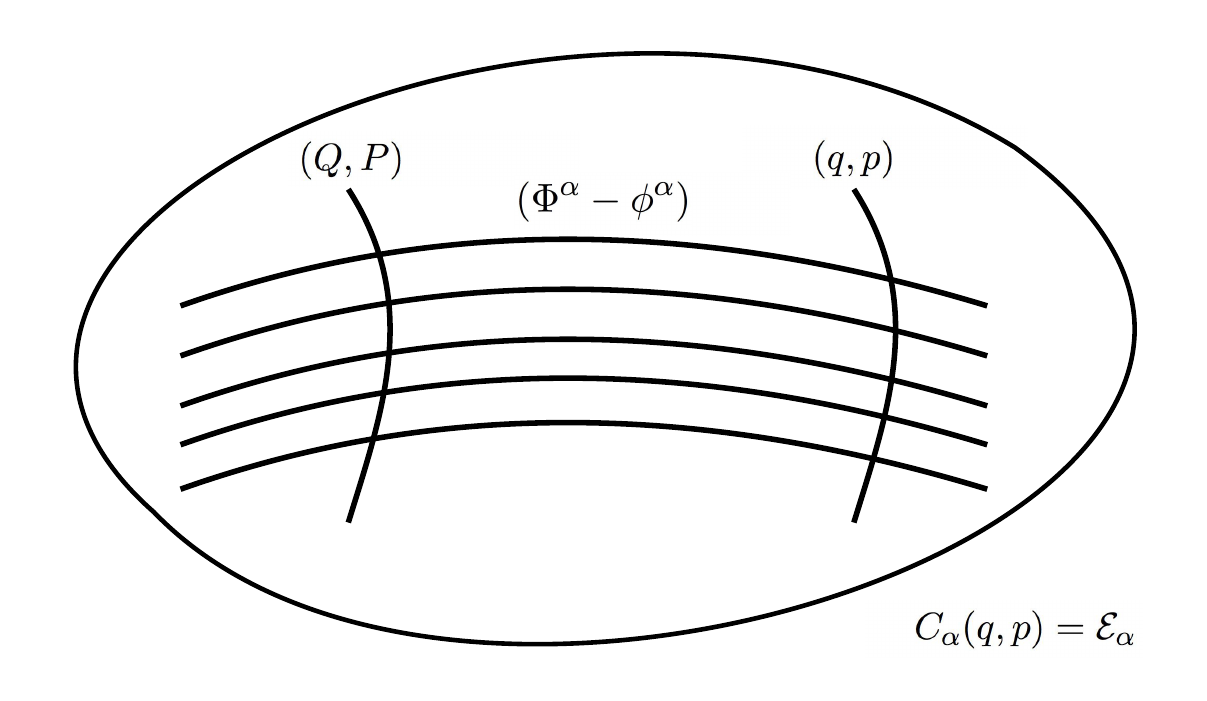}
	\end{center}
	\caption{The canonical transformation gives the general coordinates $(q,p)$ lying along the integral curves of the constraints, $C_\alpha(q,p) = \Epsilon_\alpha$, in terms of the `initial data' $(Q,P)$ and the parameters $\phi^\alpha$.  } \label{fig:can trans}
\end{figure}

We see from this that $S(q,Q; \phi^\alpha, \Phi_\alpha)$ is a type-1 generating functional for a canonical transformation taking the lower case coordinates to upper case ones subject to our requirements. Figure~\ref{fig:can trans} shows how to interpret it in terms of a flow along the integral curves of $C_\alpha(q,p)$.

It is convenient to convert $S$ to a type-2 generating functional $F(q,P; \phi^\alpha, \Phi_\alpha)$ in the $(q,p)$ coordinates by adding the boundary term $Q\wedge P$. Thus,
\begin{equation}
	\de F(q,P; \phi^\alpha, \Phi_\alpha) = p\wedge \de q + C_\alpha (q,p) \wedge \de \phi^\alpha + Q \wedge \de P\,.
\end{equation}
From the above, we find the equations defining the canonical transformation we are looking for:
\begin{subequations}
\begin{align}
	\diby F q &= p & \diby F {\phi^\alpha} &= C_\alpha(q,p) \label{eq:HJ eoms1} \\
	\diby F P &= Q  & \diby F {\Phi_\alpha} &= 0\,. \label{eq:HJ eoms2}
\end{align}
\end{subequations}
The first equation in \eqref{eq:HJ eoms1} is just the definition of the momenta $p$. Combined with the second equation in \eqref{eq:HJ eoms1}, we get the generalized Hamilton--Jacobi relations
\begin{equation}\label{gen HJ}
	\diby F {\phi^\alpha} = C_\alpha\lf(q,{\textstyle \diby F q}\rt).
\end{equation}
The second equation of \eqref{eq:HJ eoms2} simply says that, because $(Q,P)$ are preserved along the flow of the transformed constraints, there is no parameter $\Phi^\alpha$ to parametrize their flow. The first equation of \eqref{eq:HJ eoms2} becomes the Hamilton--Jacobi equation of motion.

We can reduce these equations using the following separation Ansatz for $F$
\begin{equation}\label{HJ sep1}
	F(q,P; \phi^\alpha, \Epsilon_\alpha) = W(q,P) + \Epsilon_\alpha \phi^\alpha\,,
\end{equation}
where we have slightly abused notation using the same symbol for the separations constants $\Epsilon_\alpha$ as the canonical coordinates $\Epsilon^\alpha$. This abuse is forgivable since they are equal on-shell. Our Ansatz converts the Hamilton--Jacobi relations \eqref{gen HJ} to the reduced set of equations
\begin{equation}\label{HJ red1}
	\Epsilon_\alpha = C_\alpha\lf(q,{\textstyle \diby W q}\rt).
\end{equation}
Inserting this back into \eqref{HJ sep1}, we obtain
\begin{equation}\label{main ansatz}
	F(q, P, \phi^\alpha) = W(q,P) + \phi^\alpha C_\alpha\lf(q, {\textstyle \diby W q}\rt) \,,
\end{equation}
which gives the generator of the canonical transformation we are looking for parametrized, as advertised, by $\phi^\alpha$. Using this, we can compute the final form of the Hamilton--Jacobi equation of motion
\begin{equation}\label{HJ eom}
	 \lf. Q = \diby F P \rt|_{C_\alpha\lf(q, {\textstyle \diby W q}\rt) = \Epsilon_\alpha}\,,
\end{equation}
which should be read as an equation for $q$ in terms of the `initial data' (or reference section), $(Q,P)$, the constants of motion, $\Epsilon_\alpha$, and the parameters $\phi^\alpha$. In this form, it represents an integral of motion for the $q(\phi^\alpha)$ in terms of some initial data.

An alternative formal expression can be given for the integrals of motion of the system in terms of the flow parameters $\phi^\alpha$ by directly integrating Hamilton's equations for the extended action in \eqref{eq:extended S}. These are the flow equations of the Hamilton vector field of the Hamiltonian. For the extended variables, they are
\begin{align}
	\dot \phi^\alpha &= \pb{\phi^\alpha}{H} = \lambda^\alpha & \dot \Epsilon_\alpha = \pb{\Epsilon_\alpha}{H} = 0\,.
\end{align}
Using these, we can rewrite the flow equations for $(q,p)$ as
\begin{align}
	\dot q &= \pb q H = \lambda^\alpha \pb q {C_\alpha} = \dot \phi^\alpha \pb q {C_\alpha} \\
	\dot p &= \pb p H = \lambda^\alpha \pb p {C_\alpha} = \dot \phi^\alpha \pb p {C_\alpha} \,, 
\end{align}
or
\begin{align}\label{eq:local obs}
	\delta q &= \delta \phi^\alpha \pb q {C_\alpha} & \delta p &= \delta \phi^\alpha \pb p {C_\alpha} \,.
\end{align}
We can exponentiate the Lie flow defined by these expressions as:
\begin{subequations}\label{eq:bianca obs}
\begin{align}
	q(\phi^\alpha) &= \lf. \exp \lf[ \phi^\alpha \mathcal{L}_{\mathcal{X}_{C_\alpha}} \rt] \rt. \cdot Q   = \sum_{n=0}^{n=\infty} \frac{(\phi^\alpha)^{n}}{n!}\{C_{\alpha},Q\}_n \\
	p(\phi^\alpha) &= \lf. \exp \lf[ \phi^\alpha \mathcal{L}_{\mathcal{X}_{C_\alpha}} \rt] \rt. \cdot P   = \sum_{n=0}^{n=\infty} \frac{(\phi^\alpha)^{n}}{n!}\{C_{\alpha},P\}_n \,.
\end{align}
\end{subequations}

%\begin{subequations}\label{eq:bianca obs}
%\begin{align}
%	q(\phi^\alpha) &= \lf. \exp \lf[ \phi^\alpha \pb q {C_\alpha} \rt] \rt\lvert_{q = Q} Q \\
%	p(\phi^\alpha) &= \lf. \exp \lf[ \phi^\alpha \pb p {C_\alpha} \rt] \rt\lvert_{p = P} P\,.
%\end{align}
%\end{subequations}

Where $\{C_\alpha,Q\}_0:=Q$ and $\{C_\alpha,Q\}_{n+1}=\{C_\alpha,\{C_\alpha,Q\}_{n}\}$. Here $Q$ and $P$ are required to satisfy the initial value constraint $C_\alpha(Q,P)= \Epsilon_\alpha$. This form of the integrals of motion is equivalent to solving \eqref{HJ eom}, since both result from the solution of the same variational principle. The form \eqref{eq:bianca obs} will later allow us to relate our formalism to the complete and partial  observables approach. In the quantum formalism, \eqref{HJ eom} will be analogous to a generalized Schr\"odinger formalism, while \eqref{eq:bianca obs} will correspond to a generalized Heisenberg formalism.

The interpretation of the solutions of \eqref{HJ eom} or \eqref{eq:bianca obs} depends crucially upon the difference between $\gone$ and $\gtwo$ symmetries. Although at the classical level these differences do not have empirical consequences, in the quantum context they lead to different prescriptions for the construction of the physical Hilbert space states and observables (see \S\ref{sec:rel quant}). For this reason, we will  distinguish explicitly between $\gone$ symmetries and $\gtwo$ symmetries by using different notation for the associated constraints and conserved charges. For $\gone$ symmetries, we will use the subscript $\beta$. For $\gtwo$ symmetries we will use the subscript $\mu$. If we assume that $s$ of the $r$ constraints are associated with $\gone$ symmetries, we can re-write the constraints $C_\alpha$, $\alpha=1,\hdots, r$ as $( C_\beta, C_\mu )$, $\beta=1,\hdots, s$ $\mu=s+1,...r$. We can then define $(\phi^\beta,\phi^\mu)$ and $(\Epsilon_\beta,\Epsilon_\mu)$ accordingly.

Let us consider $\gone$ symmetries first. For these symmetries, the $\phi^\beta$ parametrize unphysical flow. In such circumstances, one must impose upon \eqref{HJ eom} the additional restriction $\Epsilon_\beta = 0$. This will enforce the original constraints $C_\beta = 0$ as the generators of unphysical transformations between states. The condition $\Epsilon_\beta = 0$ can be equivalently applied to \eqref{eq:bianca obs} in terms of a restriction of the initial values of $Q$ and $P$. There are then two options\footnote{For explicit discussion of these two options see \cite[pp.113-114]{Rovelli:2004} and \cite[\S3.2]{gryb:2011}.} for the construction of observables in the case of a $\gone$ symmetry, both of which lead to Dirac observables:
\begin{itemize}
\item [i)] Retain the parameters $\phi^\beta$ and solve  \eqref{HJ eom} (or equivalently \eqref{eq:bianca obs}) for the integrals of motion, $q$, as a functions of $\phi^\beta$ and some reference section as specified by $(Q,P)$. This will give different representations of the observables  in different gauges labelled by the $\phi^\beta$. 
\item [ii)] Solve \eqref{eq:bianca obs} (or equivalently \eqref{HJ eom}) to get an explicit expression for the integrals of motion, $q$, as functions of $\phi^\beta$, $Q$ and $P$ (as per i). Invert $s$ of these expressions for the independent parameters, $\phi^{\beta}$, in terms of some subset of the integrals of motion for $q^{\beta}$ and the $(Q,P)$. Then, reinsert this result into the remaining equations for the $q$'s in terms of the $q^{\beta}$ and the full set of $Q$'s and $P$'s. Observables are then specified as families of functions $F(Q,P,q^{\beta})_{\mid q^{\beta}=\kappa^{\beta}}$ where $\kappa^{\beta}$ are a set of real numbers that pick out one member of each family of observables.  
\end{itemize}

The first strategy corresponds to a very general gauge-fixing-type methodology for constructing Dirac observables. The second corresponds to an implementation of the partial and complete observable approach \cite{Rovelli:2002,Rovelli:2004,Dittrich:2006,Dittrich:2007,Thiemann:2007} with $Q$ and $P$ the partial observables, and  $F(Q,P,q^{\beta})_{\mid q^{\beta}=\kappa^{\beta}}$ the complete observables. The first strategy has a distinct advantage over the second in that it does not depend upon the invertibility of the integrals motion, which is not generally guaranteed. Nevertheless, when they are well defined, the complete observables will be Dirac observables, and will, for $\gone$ symmetries, faithful parameterize all the physical degrees of freedom. The partial observables, on the other hand, can have different values when evaluated at phase space points corresponding to physically identical states. This means that they should not be associated with physical measurements. For $\gone$ symmetries, the Thiemann view that partial observables are non-measurable is thus supported against the Rovelli view that they can be associated with measurements. We will find further support for this conclusion concerning $\gone$ symmetries in the context of the model of \S\ref{sec:simple model}.\footnote{See \cite[p.78]{Thiemann:2007} and  \cite{Rovelli:2007,Rovelli:2014} for further discussion on this point.} 

Let us next consider the case of $\gtwo$ symmetries. In our view, these symmetries are markedly different to $\gone$ symmetries since the associated labelling parameters and conserved charges have physical significance. For the conserved charges, $\Epsilon_\mu$, this physical significance is in terms of their role as conserved quantities within the theory that have previously been interpreted as couplings of the theory (see \S\ref{sec:ambiguity} for more details). For the labelling parameters, $\phi^\mu$, the physical significance is more subtle, and is in terms of the \textit{labelling of physically distinct states of the system}. Although, in these senses, the $\Epsilon_\mu$ and $\phi^\mu$ are physically significant, they do not correspond to independent degrees of freedom. Rather they are independent parameters that parameterize physical differences between states. In order to play this role, $\Epsilon_\mu$ and $\phi^\mu$ must be retained within the classical and quantum formalisms without an increase in the overall degrees of freedom. Classically at least, this is easily achieved: although we have extended the original phase space of the theory, the presence of the first class constraint $C_\mu(q,p) - \Epsilon_\mu = 0$ removes the two extra  degrees of freedom, meaning that the total number of degrees of freedom is identical to the original un-extended phase space. Since the conserved charges are physically significant (in the sense described in \S\ref{sec:ambiguity}), no additional restrictions $\Epsilon_\mu = 0$ should be imposed upon the integral of motion given by \eqref{HJ eom}. We understand the canonical transformation generated by $F$ as reshuffling the degrees of freedom such that $\phi^\mu$ parametrizes a coordinate along the orbits of the constraints $C_\mu$. Thus, the coordinates along the orbits associated with $\gtwo$ symmetries are taken to have physical meaning: they are not `gauge' orbits. We will argue in the next section that this is precisely the case for reparametrization invariant theories, the key exemplar of a theory with $\gtwo$ symmetry. In this case, we have a single constraint  $C$, equivalent to the Hamiltonian of the theory. The Hamiltonian, of course, generates time evolution, so the $\phi$ associated with the Hamiltonian constraint parametrizes time evolution and the constant of motion $\Epsilon$ is the the total energy.\footnote{In Jacobi's action principle, which leads to a Hamiltonian constraint, the energy is considered a coupling of the theory in agreement with our considerations in \S\ref{sec:ambiguity}.}

Observables for theories with $\gtwo$ symmetries are given by the full set of $Q$s and $P$s, with the integrals motion given by solutions of \eqref{HJ eom} or \eqref{eq:bianca obs} describing genuine evolution in terms of the independent parameters $\phi^{\mu}$. Given these expressions, one has the option to apply the complete and partial observables scheme, which is equivalent to option ii) above. This will serve to eliminate the independent parameters $\phi^{\mu}$ and give an expression for the relevant Dirac observables. However,  such a construction does not give the full set of observables of the theory since there exist measurable quantities which are not described within any single representation of the complete observables. In a system with two configuration space dimensions (i.e., four phase space dimensions) and a single $\gtwo$ symmetry, for example, any construction of the complete observables will represent only half the phase space degrees of freedom, despite there being no redundancy within the theory at the level of states. The partial observables, on the other hand, will parametrize the full phase space, and can in fact be understood as the full set of observables of the theory. Moreover, since there are no `gauge orbits' on phase space, the curves defined by solving \eqref{HJ eom} or \eqref{eq:bianca obs} will be unique up to parametrization.  For $\gtwo$ symmetries, the Rovelli view that partial observables can be associated with measurements is thus supported against the Thiemann view that partial observables are non-measurable. We will find further support for this conclusion concerning $\gtwo$ symmetries in the context of the model of \S\ref{sec:simple model}. 

In our scheme, although observables are required to be constant along the flow associated with $\gone$ generating constraints (mirroring the usual Dirac requirement), no such requirement is imposed for the $\gtwo$ generating constraints. We will call the class of observables \textit{mutables}. In general, the Dirac observables will be a subset of the mutables. In particular, the Dirac  observables can be understood as the subset in which the $\Epsilon_\mu$ associated with $\gtwo$ constraints is also set to zero and method i) or ii) applied. In the following section, we will provide an explicit application of the generalized Hamilton--Jacobi construction of the mutables and also construct the relevant Dirac observables via the complete observables route. This will then lead naturally into the key physical differences that emerge in the quantum regime --- to be discussed in \S\ref{sec:rel quant}.

\section{Classical Model with Global Reparametrization Symmetry}\label{sec:simple model}

A global time reparametrization transformation induces a mapping between different parametrizations of otherwise identical configuration space curves. Formally, such symmetries can be characterized in terms of the action of the one dimensional real diffeomorphism group, $\mathcal{D}iff(\mathbb{R})$, upon the space of parametrized configuration space curves, $\mathcal{C}_{p}\in \{\gamma,t\}$, made up of a pairing of a curve, $\gamma$, and a monotonic parameter, $t$. For the case of a global time parameter $t$, the action of the reparametrization group, $\mathcal{R}ep$, takes the simple form:
\begin{equation}\label{eq:repdef}
\{ \gamma,t \} \rightarrow \{ \gamma,f(t) \}
\end{equation}
for $\frac{\de f(t)}{\de t}>0$. By definition a global reparametrization invariant theory is one in which the action is invariant under $\mathcal{R}ep$. Global reparametrization invaraince thus requires a Lagrange density that is homogeneous of order $1$ in velocities -- i.e. we need to have that:
  \begin{equation}
I=\int_\gamma L(q,\dot{q})\, \de t \rightarrow \int_\gamma \frac{\de t}{\de f(t)} L(q,\dot{q})\, \de f(t)=I\,.
\end{equation}
Crucially, this implies that the momenta will be  invariant under the action of $\mathcal{R}ep$:
\begin{equation}
p=\frac{\partial L(q,\dot{q})}{\partial \dot{q}} \rightarrow \frac{\partial L(q,\dot{q})}{\partial \dot{q}} \lf(\frac{\de f(t)}{\de t}\rt)^{-1} \frac{\de f(t)}{\de t}=p\,.
\end{equation}
Thus, instantaneous states, which are defined in terms of points in phase space, are invariant under reparametrization. Reparametrization is, therefore, a $\gtwo$ symmetry since it relates physically indistinguishable histories and there is no corresponding state symmetry relating physically indistinguishable states. Reparametrization symmetry \textit{does not} result from the existence of unphysical directions in the state space. Rather, it results purely from an invariance of the action.

Nevertheless, it is easy to see that reparametrization invariance \textit{will} always be associated with a canonical constraint. Since the Lagrange density is homogeneous of order $1$ in the velocities, Euler's homogeneous function theorem implies that the Hamiltonian density must vanish; or, rather, that it be proportional to a constraint \cite{ dirac:lectures}.  Thus, there is a precise formal basis for the existence of (global) Hamiltonian constraints in (global) reparamterization invariant theories. Such constraints have \textit{nothing whatsoever} to do with a transformation between physically indistinguishable instantaneous states. Rather, when considered as acting on phase space points, (global) Hamiltonian constraints have the unambiguous role of producing dynamical transformations between physically distinct states. It is of course true that these transformations are parametrized by arbitrary time labels, however, the image of the phase space curves that are traced out are  invariant under reparametrization of the time labels. Thus, we always have an unambiguous sequence of physically distinct phase space points, irrespective of any arbitrariness regarding the time labelling of this sequence. The existence of (global) Hamiltonian constraints is entirely consistent with $\gtwo$ symmetries.\footnote{We should note that in the context of canonical general relativity the local Hamiltonian constraints are associated with local time reparametrizations or refoliations which \textit{do} lead to nontrivial transformations of the image of phase space curves. However, this connection can only be established if one considers time-dependent smearings that trace out complete histories. Thus, it is also inappropriate to interpret local Hamiltonian constraints, in and of themselves, as generating transformation between physically indistinguishable instantaneous states.}
   
With these considerations in mind, let us consider a simple theory with both reparametrization $\gtwo$ symmetries and  the more typical $\gone$ symmetries.\footnote{An alternative example to consider would be the string worldsheet, where spatial diffeomorphisms on the worldsheet would represent a $\gone$ symmetry and time reparametrizations on the worldsheet would represent a $\gtwo$ symmetry. Indeed it is well-known that these constraints should be considered differently. It would be interesting to compare what is done in this case with what is suggested by relational quantization. } The finite dimensional particle model we will study has a total Hamiltonian given by:
\begin{equation}
	H(\vec q_i, \vec p_i) = N \ham (\vec q_i, \vec p_i) + \vec \lambda \cdot \vec \gau(\vec p_i),
\end{equation}
which is defined on the phase space $\Gamma(\vec q_i, \vec p_i)$, where $i$ ranges over the number of particles $n$. In this free `Jacobi' theory, evolution of the particle positions, $\vec q_i$, and momenta, $\vec p_i$, is generated by the Hamiltonian constraint
\begin{equation}\label{eq:model ham}
	\ham(\vec q_i, \vec p_i) = \sum_i \frac{ \vec p_i^2 }{2m_i} - E \approx 0\,,
\end{equation}
which characterizes a $\gtwo$ symmetry, as we've just explained. The `Gauss-like' constraint,
\begin{equation}\label{eq:model P}
	\vec \gau = \sum_i \vec p_i \approx 0\,,
\end{equation}
implements a gauging of the spatial translational invariance of the free particle system. The lapse function $N$ and the vector $\vec \lambda$ act as Lagrange multipliers enforcing the vanishing of these constraints. The vanishing of the total linear momentum implies that the position of the centre of mass, ${\vec q}_\text{cm} = \frac 1 {\sum_i m_i} \sum_i {\vec q}_i$, is pure gauge. This gauge symmetry is interpreted in terms of a closed system that can only measure inter-particle separations and has no way of physically distinguishing between one value of its centre of mass and another. It is, therefore, characterized unambiguously as a $\gone$ symmetry.

This model is particularly useful for understanding canonical general relativity whose constraints also split into an evolution generator (the ADM Hamiltonian constraint) and a generator of an instantaneous symmetry (the ADM Diffeomorphism constraint). It is inspired by more general Barbour--Bertotti models which have been studied extensively \cite{ barbourbertotti:mach, anderson:found_part_dyn, anderson:rel_part_mech_1, anderson:rel_part_mech_2}.

Following the general procedure outlined in \S \ref{sec:general formalism}, we define the extended theory
\begin{equation}
	S_\text{e} = \int \de t \lf[  \sum_i \dot{\vec q_i} \cdot \vec p_i + \dot{\vec{\sigma}}\cdot \vec{\Upsilon} - \dot \tau \Epsilon - N \lf( \Epsilon - \ham \rt) - \vec \lambda \cdot \lf( \vec{\Upsilon} - \vec\gau \rt) \rt]\,,
\end{equation} 
where the extended variables $(\tau, \vec{\sigma} ) $ are arbitrary labels parametrizing the time and centre of mass of the system respectively. The energy, $\Epsilon$, can be thought of as a redefinition of the zero of the total energy of the system $E\to E + \Epsilon$. The other conjugate momentum variable, $\vec{\Upsilon}$, is the total linear moment of the system. We have conventionally added a minus sign to the $\dot\tau \Epsilon$ term to ensure the usual relations between time and energy. The extended theory is physically indistinguishable from the original in the case of the $(\tau, \Epsilon)$ extension, and when $\vec{\Upsilon} = 0$ in the case of the $(\vec{\sigma}, \vec{\Upsilon}) $ extension.

After making the appropriate identifications, the Ansatz \eqref{main ansatz} for $F$, takes the form
\begin{equation}\label{free Jac ansatz}
	F(\vec q_i, \vec P_i; \vec{\sigma} , \tau) = W(\vec q_i, \vec P_i) + \vec \gau\lf(\vec q_i, \sfrac{\partial W}{\partial\vec q_i} \rt) \cdot \vec \sigma - \ham\lf(\vec q_i, \sfrac{\partial W}{\partial \vec q_i} \rt)\, \tau \,,
\end{equation}
where $\ham$ and $\vec \gau$ obey the reduced Hamilton--Jacobi relations
\begin{align}\label{eq:model red HJ}
	\Epsilon &= \ham\lf(\vec q_i, \sfrac{\partial W}{ \partial\vec q_i} \rt) & \vec {\Upsilon} &= \vec \gau\lf(\vec q_i, \sfrac{\partial W}{ \partial \vec q_i} \rt)\,.
\end{align}
Using the definitions \eqref{eq:model ham} and \eqref{eq:model P} and the additional Ansatz for $W$
\begin{equation}\label{free Jac W ansatz}
	W(\vec q_i,\vec P_i) = \sum_i \vec q_i \cdot \vec P_i,
\end{equation}
we get
\begin{equation}
	F(\vec q_i, \vec P_i; \vec {\sigma} , \tau) = \sum_i \vec q_i \cdot \vec P_i - \lf( \sum_i \frac {\vec P_i^2} {2m_i} - E \rt) \tau + \vec {\sigma} \cdot \sum_i \vec P_i\,.
\end{equation}
The Hamilton--Jacobi equation of motion, \eqref{HJ eom}, for this system is then
\begin{equation}
	\diby F {\vec P_i} = \lf( \vec q_i + \vec {\sigma}\rt) - \frac {\vec P_i \tau} {m_i} = \vec Q_i,
\end{equation}
which can easily be inverted for $\vec q_i$
\begin{equation}\label{eq model iom}
	\vec q_i = \lf( \vec Q_i - \vec {\sigma} \rt) + \frac {\vec P_i \tau} {m_i}\,.
\end{equation}
The $\vec P_i$'s in this equation must obey the constraints
\begin{align}\label{eq:model P constraints}
	\Epsilon &= \sum_i \frac {\vec P_i^2} {2m_i} - E & \vec \gau &= \sum_i \vec P_i = 0
\end{align}
in order for the reduced Hamilton--Jacobi equations \eqref{eq:model red HJ} to be satisfied. Note crucially that, while imposing $\vec \gau = 0$ is necessary for the classical equivalence of the extended and un-extended theories, no such condition is required on $\Epsilon$, which only contributes a shift of the total energy (more on the interpretation of this in \S\ref{sec:ambiguity}). These last equations correspond simply to constraints on the freely specifiable initial momenta. 

This theory reproduces precisely our expectations: it gives the usual integral of motion for the free particle plus an extra term that just shifts the origin of each particle system by $\vec {\sigma}$, which is clearly the coordinate along the gauge orbit for this theory. Exploring the various different ways of interpreting the integral of motion \eqref{eq model iom} gives insight into the nature of observables and how they differ for $\gone$ and $\gtwo$ symmetries. 

Consider first the $\gone$ symmetry associated with the coordinate $\sigma$. For this simple symmetry, there is a natural way to parametrize the reduced phase space. If we sum \eqref{eq model iom} over the mass weighted particle indices and divide by the total mass, $m_\text{tot} = \sum_i m_i$, we obtain (upon using the second constraint of \eqref{eq:model P constraints})
\begin{equation}
	\vec{\sigma} = \vec Q_\text{cm} - \vec q_\text{cm}\,,
\end{equation}
which is just the change in the centre of mass of the system. Reinserting this back into the integral of motion \eqref{eq model iom} gives
\begin{equation}\label{eq:model red iom}
	\vec q^\text{cm}_i = \vec Q^\text{cm}_i + \frac {\vec P_i}{m_i} \tau\,,
\end{equation}
where the `cm' refers to the use of centre of mass coordinates. Because of the redundancy implied by the translational symmetry, the centre of mass coordinates over-parametrize the reduced phase space. We then have the choice between options i) and ii) detailed above. In this case, i) corresponds to treating \eqref{eq:model red iom} as an over-complete set of relational observables. These are easily verified to be Dirac observables for the system. Option ii), on the other hand, leads us to treat the coordinates $\vec q_i$ as explicit functions of the initial data $(\vec Q_i, \vec P_i)$ and the change in centre of mass $\vec{\sigma}$ (and the time, $\tau$). This is equivalent to treating the change in centre of mass as a partial observable that parametrizes the different representations of the coordinates $\vec q_i$ in different gauges. Although this can be done for this system, it is very unnatural because it parametrizes things that are not physically measurable --- the explicit coordinates, $\vec q_i$ --- in terms of something else that is not physically measurable --- the change in centre of mass, $\vec{\sigma}$. For this example of $\gone$ symmetries, the Thiemann perspective, in which partial observables are non-measurable quantities, is more appropriate than the Rovelli perspective, where the partial observables are understood as measurable. Moreover, since it is only the Dirac observables that have operational significance, one could simply follow the more direct definition via option i) and avoid the partial and complete observables entirely.

Contrast this with what happens when we apply the same options to $\gtwo$ symmetries. Option i) is just equivalent to a standard gauge fixing, which we have the ability to do arbitrarily. For example, the integral of motion in terms of centre of mass coordinates is equivalent to using a gauge fixing $\vec{\sigma} = 0$. The analogous condition $\tau = 0$ indeed gives us a parametrization of the reduced phase space. Then, the `integral of motion' reduces to the trivial statement
\begin{equation}
	\vec q^\text{cm}_i = \vec Q^\text{cm}_i\,.
\end{equation}
The only non-trivial equations in this gauge are, perhaps unsurprisingly, the initial value constraints \eqref{eq:model P constraints}. Clearly this option does not provide us with any dynamical information. We will see in Section 5 that this timelessness is also found when Dirac quantization is applied to this manifestly dynamical model. This should not be surpassing since Dirac quantization rests of the assumption of $\gone$ symmetries.

The only way to obtain a notion of evolution for this system is to consider \eqref{eq:model red iom} as a genuine evolution equation for the system. Rather than taking the timeless route and following option i) towards the construction of Dirac observables for $\gtwo$ symmetries, one follows the alternative mutables route. We can identify the physically relevant observables of the system as those corresponding to the \emph{entire} set of configuration space variables $\vec q^\text{cm}_i$ after removing the centre of mass. These are the mutables for this system since they commute with the `Gauss' constraints $\gau$ but not the Hamiltonian constraint $\ham$. These mutables evolve according to \eqref{eq:model red iom}, tracing out curves labelled by the arbitrary parameter $\tau$, which is of course itself not an observable. Rather, $\tau$ is an independent parameter, and, as such, can be specified independently of quantities which are deemed measurable within the theory. The curves defined by \eqref{eq:model red iom} are reparametrization invariant even if the equation makes reference to the unphysical labelling parameter. Thus, the mutables, as we have defined them \textit{are} invariant under the relevant $\gtwo$ symmetry. 

One might, however, wish to give a `parameter free' expression for the relative variation of the observables. It is in that context that the  complete and partial observables program --- option ii) --- offers a unique strength. Consider our formalism in 1-dimension. We can choose the partial observables to be the centre of mass coordinates, $ q^\text{cm}_{i}$, meaning the $ q^\text{cm}_{i}$ defined via \eqref{eq:model red iom}, play the role of the `flow equations'. A natural choice of clock variables is the centre of mass coordinate of one of the particles, say  $q^\text{cm}_1$. We can invert \eqref{eq:model red iom} for particle 1 to obtain
\begin{equation}
	\tau = \frac {m_1}{P_1} \lf( Q^\text{cm}_1 - q^\text{cm}_1 \rt)\,.
\end{equation}
The first initial value constraint of \eqref{eq:model P constraints} gives $P_1$ in terms of the remaining $P_i$'s. Using this, we can deparametrize the evolution of the remaining $q_a$'s (where $a = 2\hdots n$) in terms of $q_1$
\begin{equation}
	q_a^\text{cm} (Q_i^\text{cm},P_i,q_1^\text{cm}) = Q_a^\text{cm} - \frac {m_1} {m_a} P_a \frac { \lf( Q_1^\text{cm} - q_1^\text{cm} \rt) } { \lf[ 2 m_1\lf( E + \Epsilon - \sum_a \frac {P_i^2} {m_i} \rt) \rt]^{1/2} }\,.
\end{equation}

For any $q_1^\text{cm}=\kappa \in\mathbb{R}$ this expression defines a `complete observable', which will also be a Dirac observable. Since these observables are defined as functions of the mutables they constitute a subset of them. In practice, then, one can use the complete and partial observables program to deparametrize the evolution purely in terms of observable quantities. This evolution is, however, fundamentally controlled by \eqref{eq:model red iom} and is \emph{always} well-defined, even when a particular deparametrization breaks down. Thus, on our view, even if one wishes to use parameter-free `complete observable' expressions, one is still required to retain the full `partial observables' representation given by \eqref{eq:model red iom}. This indicates that the the Rovelli perspective, in which partial observables are measurable quantities, is more appropriate than the Thiemann perspective, where the partial observables are understood as non-measurable. Moreover, it coincides with a scheme where the mutables, and not the Dirac observables, are fundamental.    

Our example has thus illustrated the following conclusions: 1) standard gauge fixings and parametrizations of the reduced phase space are the appropriate techniques for dealing with $\gone$ symmetries; 2) in the case of $\gtwo$ symmetries, the evolution of the system is described by the reduced integral of motion \eqref{eq:model red iom} in terms of the arbitrary parameter $\tau$, which is always well-defined; 3) in that context, there are good motivations for applying the complete and partial observables program in order to construct parameter-free complete observable expressions. However, one must retain the parameter dependent `partial observable' expressions in order to fully represent the physics of the system. In \S\ref{sec:rel quant}, we will argue that such interpretational differences between $\gone$ and $\gtwo$ symmetries at the classical level imply a physical difference at the quantum level. This difference can be characterised in terms of a different behaviour for observables in certain quantum theories, including the model constructed in this section.
 
\section{Physical Interpretation of Extended Momentum}\label{sec:ambiguity}

Given the extension procedure developed above, it is important to understand the roles played by the extended variables $(\phi^\alpha, \Epsilon_\alpha)$. For $\gone$ symmetries, this is straightforward since ultimately our treatment of them requires imposing $\Epsilon_\beta = 0$. The dynamics of the extended theory is then manifestly equivalent to the original. The $\phi^\beta$ become labels that parametrize unphysical flow on the constraint surface. 

For $\gtwo$ symmetries, our proposal is to treat the flow parametrized by $\phi^\mu$ as physical. This suggests that we put no additional restrictions on $\Epsilon_\mu$ since we wish to identify all the original configuration space variables as physical. It is then important to understand the physical significance of the unconstrained constants of motion $\Epsilon_\mu$. In general, this will depend on a detailed study of the physics of the system. Here, we will restrict to the case where the only $\gtwo$ symmetry present in the system is that of a reparametrization symmetry. In this case, the relevant $\gtwo$ constraint is a Hamiltonian constraint $C_\mu \to \ham$ and the parameter associated to its flow is the time $\phi^\mu \to \tau$. We will now see that it is always possible to associate the momentum, $\Epsilon$, with a coupling constant of the theory, which should be interpreted as the total energy.

Consider the decomposition of $\ham$ into the general form
\begin{equation}
	\ham = T(q,p) + \sum_I c^I V_I(q)\,,
\end{equation}
where $T(q,p)$ is a kinetic term, $V_I(q)$ are different potentials terms, and $c^I$ are coupling constants associated to each $V_I(q)$. The general Hamiltonian for the theory takes the form
\begin{equation}
	H = N \ham + \lambda^\beta C_\beta\,.
\end{equation}
Because the Lagrange multiplier $N$ can take any form, we can define an equivalent Hamiltonian by multiplying the Hamiltonian constraint by an \emph{arbitrary}, nowhere vanishing function, $f(q,p)$, such that $\ham^\prime = f(q,p) \ham$ provided we redefine $N$ appropriately: $N^\prime = N/f(q,p)$. By using $f(q,p) = 1/V_i(q)$ for some arbitrarily selected value of $i$, one can always find an equivalent expression for $\ham$ which singles out a particular coupling $c^i$ such that it appears as a constant in the transformed Hamiltonian. The interpretation of this form of the Hamiltonian is that, by changing the lapse, $N$, you are using a parametrization such that $c^i$ is interpreted as the total energy of the system.

This, it would seem, introduces an ambiguity for the quantization of a reparametrization invariant theory. The reason for this potential ambiguity is that, since any coupling of the theory can be interpreted as an energy for some parametrization, how should one define the vacuum state --- which is defined by the state that minimizes the energy --- in an unambiguous way? In practice, this is done by privileging a set of inertial observers who see the `true' vacuum of the system. For a harmonic oscillator potential, for example, it is clear which normalisation of the Hamiltonian constraint should be used to define the vacuum of the system: no one could confuse a spring constant for an energy. Thus, in practice, any potential ambiguity should be resolved by the physical properties of the system. In many ways, this situation is analogous of the Unruh effect in quantum field theory, which also privileges the vacuum of a class of inertial observers.

The implications of this for the current discussion follow from the observation that the extension procedure $\ham \to \ham - \Epsilon$ can be applied equivalently to \emph{any} of the normalizations of the Hamiltonian constraint. The interpretation of such a procedure is to single out one of the coupling constants $c_i$ of the theory and promote it to a constant of motion. Although this should not change the interpretation of the classical theory because it is equivalent to a redefinition of the coupling $c_i \to c_i + \Epsilon$, care must be taken with regard to the implications to the quantum theory. It would be unwise to introduce a procedure that allows, for example, a harmonic oscillator system to be in a superposition of different values of the spring constant. To avoid such a scenario, one must identify the class of inertial observers of the system in question that are used to define the `true' vacuum of the theory. This singles out the unique coupling of the theory that can be associated with the energy of the system. In turn, this fixes the normalization of the Hamiltonian constraint that is to be used to perform the extension procedure required for relational quantization, which we will now describe in detail.

\section{Relational Quantization}\label{sec:rel quant}

The most important implication of the foregoing considerations is to be found within the quantum realm. All standard quantization procedures for theories with gauge symmetry --- including path integral \cite{Faddeev:1969}, constraint \cite{Dirac:1964,Giulini:1999a}, or reduced phase space \cite{sniatycki:1983,thiemann:2006b} approaches --- are founded upon a particular identification of the physical degrees of freedom that assumes one must eliminate redundancy within the representation of the physical states. While such an identification is well justified for $\gone$ symmetries, it is not for $\gtwo$. The case of reparametrization invariance explicitly illustrates this point since, as discussed above, such a symmetry can only be understood in terms of redundancy at the level of histories, and has no implications for redundancy at the level of states. 

The requirements for a quantization technique that faithfully perseveres the physical characteristics of a classical theory with $\gtwo$ symmetries are clear. First, such a technique should take the classical mutables as the basis for the algebra of quantum observables, which are defined as Hermitian operators on a physical Hilbert space. Second, we should define quantum wavefunctions as the rays of this physical Hilbert space that are invariant under change with respect to the independent parameters of whatever $\gone$ symmetries might be present in the theory. In other words, we require that $\gone$ symmetries be treated equivalently to standard gauge theory techniques. Third, such quantum wavefunctions should evolve according to an evolution equation that reduces in the semi-classical limit to the Hamilton--Jacobi evolution equation \eqref{gen HJ} with respect to the generator of the $\gtwo$ symmetry. In particular, this should imply that the wavefunction should be able to exist in superpositions of eigenstates of the constant of motion associated with the $\gtwo$ symmetries. 

Relational quantization is a technique specifically geared to fulfil these requirements. In previous work, the articulation of relational quantization depended upon an explicit extension procedure combined with standard quantization techniques. Here, we use the generalised Hamilton--Jacobi formalism introduced above to give a more direct presentation. This treatment follows the well-trodden road from the Hamilton--Jacobi characteristic functional to the quantum mechanical wave equation (see for example \cite[pp. 99-109]{Rund:1966}). Essentially, one considers families of hypersurfaces of constant value of the characteristic function as wavefronts propagating in configuration space with respect to the independent parameters. The crucial step is then to interpret these wavefronts as surfaces of constant phase of a complex valued wavefunction on configuration space evolving with respect to the independent parameters. In the context of our formalism, this means one takes the functional $F(q,P,\phi^\alpha)$ of \eqref{main ansatz} as the basis for a complex wavefunction $\Psi(\phi^\alpha)$, defined on a Hilbert space labelled by the eigenvalues of a complete set of the mutables $(\hat q, \hat p)$ and the independent parameters $\phi^\alpha$. For the moment, we will not distinguish between $\gone$ and $\gtwo$ symmetries but will discuss the differences later.
 
As per standard canonical quantization, one takes the symplectic structure of $\Gamma$ to generalize naturally to the operator algebra $\pb q p = \mathbbm 1 \to \lf[ \hat q, \hat p\rt] = i\hbar \hat{\mathbbm 1}$.\footnote{Strictly speaking a more  rigorous algebraic quantization method is required to insure these brackets (as well as the structures below) are well defined. Such formal issues are tangential to our current project and so can reasonably be neglected. See \cite{isham:1984a,isham:1984b,henneaux_teit:quant_gauge,Giulini:1999a,Giulini:1999b,Thiemann:2007} for more details.}   The functional dependence of the wavefunction on the independent variables $\phi^\alpha$ is given by the equations: 
\begin{equation}\label{quant ev}
	C_\alpha(\hat q, \hat p) \Psi = -i\hbar \diby{\Psi}{\phi^\alpha},
\end{equation}
which are the direct generalization of \eqref{gen HJ}. We can now posit the separation Ansatz (the analogue of \eqref{HJ sep1})
\begin{equation}\label{quant sep1}
	\Psi = \sum_{n} c_{n} \exp(-i\hbar\, \Epsilon^n_\alpha \phi^\alpha) \psi_n,
\end{equation}
where $n$ labels all eigenstates of the operator $C_\alpha(\hat q, \hat p)$
\begin{equation}
	C_\alpha(\hat q, \hat p)\, \psi_n = \Epsilon^n_\alpha\, \psi_n\,.
\end{equation}
These are the analogues of the reduced Hamilton--Jacobi equations \eqref{HJ red1}. As we shall see later, these equations play a similar role within the theory to the time-independent Schr\"odinger equation.

The appropriate definition of the Hilbert spaces on which these operators act and the construction of the inner product under which observables should be self-adjoint depends upon the physical interpretation of the equations \eqref{quant ev}.

In the case of a $\gone$ symmetry, the change parameterized by the independent  parameters is unphysical. This means that \eqref{quant ev} should be interpreted as quantum flow equations parametrizing physically indistinguishable quantum states. Resultantly, one should follow the standard Dirac reasoning and impose the following \textit{kinematical} restrictions on the theory: a) physical states of the quantum theory, $\Psi_\text{phys}$, should be $\phi^\beta$-independent; and b) the physical Hilbert space should be constructed in such a way that the inner product is invariant under the action of the $C_\beta$'s. (Recall that we are again using the index $\beta$ to label $\gone$ symmetries only.)

In the case of a $\gtwo$ symmetry, the change parameterized by the independent  parameters is physical. This means that the equations \eqref{quant ev} should be treated as \textit{dynamical} equations giving the evolution of the quantum states with respect to the $\phi^\mu$. (Recall that we are again using the index $\mu$ to label $\gtwo$ symmetries only.) This implies no kinematical restriction, and so places no further requirements upon the possible physical states, physical inner product, or observables. 

Our analysis of $\gone$ symmetries coincides with the standard approach to gauge symmetries. For this case, we can make use of rigorous methodologies within the existing literature to implement conditions a) and b). One such technique is \textit{group averaging}, which is a key aspect of Refined Algebraic Quantization (RAQ) \cite{Giulini:1999a}. Essentially, one defines the physical states by averaging $\Psi$ over the group manifold. For simple cases\footnote{For more complex cases, such as where zero is not in the the discrete spectrum of the constraint operator, the full machinery of Refined Algebraic Quantization would be needed. The generality of RAQ as a technique for the quantization of theories with $\gone$ symmetries extends to all cases where: i) the constraints are self-adjoint operators with an algebra which closes with structure constants; ii) the symmetry group is a finite dimensional locally compact group with respect to a suitable topology. Notably, these conditions are not satisfied for the Hamiltonian constraints of general relativity, where other techniques are required \cite{Thiemann:2007}.} this averaging takes the form:
\begin{align}
	\Psi_\text{phys} &= \int \de\phi^\beta\, \Psi \\
					 &= \int \de\phi^\beta\, \sum_n c_n \exp(-i\hbar \Epsilon_\beta^n\, \phi_\beta ) \psi_n \\
					 &= \sum_n c_n \delta(\Epsilon^n_\beta) \psi_n \\
					 &= c_{0,\beta} \psi_{0,\beta}.					
\end{align}
The group averaging thus superselects the eigenstate of energy zero, and we recover the results of Dirac's original constraint quantization scheme: i.e. that the physical states should satisfy
\begin{equation}
	C_\beta(\hat q, \hat p) \psi_\text{phys} = 0.
\end{equation}

We can use the group averaging to define a rigging map, $\eta$, from the general states $\Psi$ to the physical states. An inner product can then naturally be defined on the physical Hilbert space using this map:
\begin{equation}
\langle \eta(\Psi^1),\eta(\Psi^2 )\rangle =
	\langle \Psi^1_\text{phys},\Psi^2_\text{phys} \rangle = \langle \psi^1_0,\psi^2_0 \rangle.
\end{equation}

The most important point here is that the construction of the physical Hilbert space involves a quantum analogue of  classical phase space reduction. The rigging map is a projection which removes unphysical states, but it also has a further role in `quotienting out' physically identical states. These states lie upon a `quantum gauge orbit' of parametrized by $\phi^\beta $: $\Psi(\phi^\beta) = e^{ \phi^\beta \hat C_\beta }  \Psi(0)$ \cite{Corichi:2008}. This is the direct integration of the quantum flow equation \eqref{quant ev} for the $\gone$ symmetries. Any two points along this orbit yield the same physical state and so, in group averaging, we are explicitly removing kinematical redundancy at the level of states. This is true both in simple cases of $\gone$ symmetries, such as those considered here, or more generally when such symmetries are treated within Refined Algebraic Quantization. We can see this in the latter case explicitly since one variant of the RAQ definition of the physical Hilbert space involves explicitly taking the quotient of the kernel of the rigging map \cite{Giulini:1999a}. We thus again see why application of standard gauge quantization techniques to $\gtwo$ symmetries is inappropriate: such techniques assume redundancy at the level of states not histories. Applying a `quantum quotienting operation' to $\gtwo$ symmetries amounts to identifying physically distinct states corresponding to different values of the evolution parameters $\phi^\mu$. This is to confuse dynamics with kinematics, and change with redundancy. 

Once the definition of the physical Hilbert space is given, observables can be constructed by looking for operators that are self-adjoint with respect to this inner product. The kinematic restrictions that we have imposed with regard to $\gone$ symmetries but not $\gtwo$ symmetries ensure that such observables will be quantum mutables. Of particular note is the (generalised) Heisenberg picture construction of the mutables. Since the equations \eqref{quant ev} are dynamical equations, not kinematic restrictions, we can consider a Heisenberg picture with the physical change in observables parametrized by the relevant independent parameters $\phi^\mu$. This leads to the Heisenberg evolution equations for the mutables $\hat {\mathcal O} $:
\begin{equation}\label{eq:Heisenberg}
	i \hbar \frac {\de}{\de \phi^\mu} \hat {\mathcal O} = [\hat {\mathcal O}, \hat C^\mu]\,,
\end{equation}
which are the quantum analogues of \eqref{eq:local obs}. These exponentiate to the well-known expressions
\begin{equation}
	\hat {\mathcal O}(\phi^\mu) = e^{-i\hbar\phi^\mu \hat C_\mu^\dag} \hat {\mathcal O}(0) e^{i\hbar\phi^\mu \hat C_\mu} \,,
\end{equation}
which are the analogues of \eqref{eq:bianca obs}.

The above considerations make clear the main difference between relationally quantized theories and Dirac quantized theories. In relational quantization, there is a larger set of observables because these are not required to commute with the $\gtwo$ constraints. These observables evolve according to the Heisenberg equations \eqref{eq:Heisenberg}. This implies that we have a different quantum state than that obtained through Dirac quantization. Specifically, in relational quantization the state is allowed to be in a general superposition of eigenstates of the evolution operators $C^\mu$ according to \eqref{quant sep1}. While in the classical theory the introduction of the extended variables $(\phi^\mu, \Epsilon_\mu)$ did not change the physical predictions of the theory, the same is no longer true at the quantum level: relational quantization leads to physically different predictions for the behaviour of observables as compared with Dirac quantization.

Another way to see this difference is to note that option ii) of \S\ref{sec:general formalism} is no longer available. It is not possible to simply single out one of the mutables, $\hat{\mathcal O}$, and invert its complete integral in terms of the evolution parameters $\phi^\mu$. This is for the simple reason that the observables are now operators, and operator equations can not be solved in terms of parameters. Instead, we can only apply the quantum analogue of option ii) in a dynamical limit of the theory where the Heisenberg state of system is such that some observables $\hat{\mathcal O}$ is highly peaked around its expectation value. In this case, one can approximate the evolution of $\hat{\mathcal O}$ by taking expectations values of both sides of \eqref{eq:Heisenberg} (formally, this approximation involves ignoring the higher moments of $\mean{\hat{\mathcal O}}$). Then it is at least possible, although not guaranteed, that this expression can be inverted for $\phi^\mu$ in terms of $\mean{\hat{\mathcal O}}$. Thus, only in such a limit is it possible to recover something like the quantum analogue of the complete and partial observables framework.

In this way, not only do we recover a quantum version of the complete and partial observables program, but our approach provides a more general framework for understanding what happens when this program breaks down. From the point of view of relational quantization, non-unitary evolution of the complete observables in terms of a particular partial observable occurs when the semi-classical approximation required to treat that partial observable as a clock breaks down. In our approach, the evolution is always unitary and well-defined provided the physical inner product can be constructed in such a way that the evolution operators $\hat C_\mu$ are self-adjoint (this is necessary requirement to being able to apply our procedure consistently).

\section{Simple Quantum Model}

Application of relational quantization to the model of \S\ref{sec:simple model} is as follows. The quantum version of the generalized Hamilton--Jacobi relations \eqref{gen HJ} reads
\begin{align}
	-i\hbar \diby {\Psi(\tau, \vec \sigma)} \tau &= \lf( \sum_i \frac {\hat{\vec {p}}_i}  {2m_i}  - E \rt) \Psi(\tau,\vec \sigma)& i\hbar\diby {\Psi(\tau,\vec \sigma)} {\vec \sigma}  &= \lf( \sum_i \hat{\vec{p}}_i \rt) \Psi(\tau, \vec \sigma)\, ,
\end{align}
for the wavefunction $\Psi(\tau,\vec \sigma)$, which represents a state in the auxiliary Hilbert space and is dependent upon both the independent parameters $\tau$ and $\vec \sigma$.

The Ansatz \eqref{free Jac ansatz} generalizes to
\begin{equation}
	\Psi(\tau, \vec \sigma) = \int \de \Epsilon\, \de^3 \vec{\Upsilon}\, c(\Epsilon,\vec{\Upsilon}) \exp \lf\{ i \hbar \lf( \Epsilon \tau + \vec{\Upsilon} \cdot \vec \sigma  \rt)\rt\} \psi_{\Epsilon,\vec{\Upsilon}}\,.
\end{equation}
This leads to the reduced equations
\begin{align}\label{eq:quantum_red}
	\lf( \sum_i \frac {\hat{\vec {p}}_i^{\, 2}}  {2m_i}  - E \rt) \psi_{\Epsilon,\vec{\Upsilon}} &= \Epsilon \psi_{\Epsilon,\vec{\Upsilon}} & \lf( \sum_i \hat{\vec{p}}_i \rt) \psi_{\Epsilon,\vec{\Upsilon}} &= \vec{\Upsilon} \psi_{\Epsilon,\vec{\Upsilon}}\,,
\end{align}
which are the analogues of \eqref{eq:model red HJ}. The $\psi_{\Epsilon,\vec{\Upsilon}}$ are thus energy and total momentum eigenfunctions. These eigenfunctions and their eigenvalues can be constructed in analogy to the Ansatz \eqref{free Jac W ansatz}
\begin{equation}\label{eq:eigenstates}
	\psi_{\Epsilon,\vec{\Upsilon}} = \exp \lf( -i\hbar \sum_i \vec P_i(\Epsilon, \vec \Upsilon) \cdot \vec q_i \rt)\,.
\end{equation}
Here, we have used a configuration basis where
\begin{align}
	\hat{\vec q}_i &= \vec q_i & \hat{\vec p}_i & = -i\hbar \diby{}{\vec q_i}\,
\end{align}
and the $P_i(\Epsilon, \vec \Upsilon)$ are integration constants depending on $(\Epsilon, \vec \Upsilon)$. Their dependence on $(\Epsilon, \vec \Upsilon)$ is obtained implicitly by the fact that they must obey the relations
\begin{align}\label{eq:quantum const}
	\Epsilon + E &= \sum_i \frac {\vec{P}^{\, 2}_i}{2m_i} &  \vec{\Upsilon}&= \sum_i \vec P_i\,,
\end{align}
which are obtained by plugging the Ansatz \eqref{eq:eigenstates} into the reduced equations \eqref{eq:quantum_red}. Note that the states \eqref{eq:eigenstates} are all non-normalizable, as is usually the case for the free particle, so that the usual care (i.e., some regularization procedure) must be taken to define an inner product on states.

For the Hamiltonian constraint, no group averaging is required because change parameterized by $\tau$ is interpreted as being physical. However, for the total momentum constraint, the transformations parameterized by the $\vec \sigma$ are manifestly unphysical. They are simply `Leibniz shifts' expressed as redefinitions of the origin of the centre of mass. Thus, group averaging is required in this case. For this system, this involves constructing the physical states
\begin{align}
	\Psi_\text{phys}(\tau) &= \int \de^3 \vec \sigma \Psi(\tau, \vec \sigma) \nonumber \\
		 &= \int \de \Epsilon\, c(\Epsilon,0) \exp \lf\{ i\hbar \lf( \Epsilon \tau - \sum_i \vec P_i \cdot \vec q_i \rt) \rt\} \label{eq:model soln}
\end{align}
where, in the last line, the integration over both $\vec\sigma$ and $\vec\Upsilon$ simply sets $\vec\Upsilon = 0$ (note that the $P_i(\Epsilon, \vec \Upsilon)$ still must satisfy \eqref{eq:quantum const}). The inner product on physical states can now similarly be defined through the relevant rigging map. 

The mutables are all operators constructed from the centre of mass operators $\hat {\vec q}^\text{ cm}_i = \hat{\vec q}_i - \frac 1 {M_\text{tot}} \sum_i \frac {\hat{\vec q}_i}{m_i} $ (where $M_\text{tot}$ is the total mass) and the conjugate operators $\hat {\vec p}^\text{ cm}_i \equiv \hat{\vec p}_i - \frac 1 n \sum_i \hat{\vec p}_i$. Their evolution in terms of $\tau$ is given by the Heisenberg evolution equations
\begin{equation}
	-i\hbar \frac {\de}{\de\tau}\hat{\mathcal O}(\hat {\vec q}^\text{\,cm}_i,\hat {\vec p}^\text{\,cm}_i, \tau) = [\hat{\mathcal O}, \hat \ham]\,
\end{equation}
where
\begin{equation}
	\hat \ham = \sum_i \frac {\lf(\hat{\vec {p}}_i^\text{ cm}\rt)^2}  {2m_i}  - E\,.
\end{equation}
These clearly exceed the number of Dirac observables, which must also commute with the Hamiltonian constraint $\hat\ham$ and, thus, can have no evolution with respect to it. Furthermore, the wavefunction, $\Psi$, of this theory can be formed from an arbitrary superposition of energy eigenstates according to \eqref{eq:model soln}. Note that these states \emph{can} be made normalizable unlike the pure states one obtains from Dirac quantization.

\section{Conclusion}

In this paper, we first isolated the important physical distinction between $\gone$ gauge symmetries, which exist at the level of histories and states, and $\gtwo$ gauge symmetries, which exist at the level of histories and not states. This distinction was characterised explicitly using a generalized Hamilton--Jacobi formalism within which a non-standard prescription for the observables of classical totally constrained systems was developed. These ideas were then used to motivate a `relational quantisation' procedure which is different from Dirac quantization. In particular, relational quantization of totally constrained systems was shown to lead to a formalism that allows for superpositions of energy eigenstates and includes an enlarged set of quantum observables: the `mutables'. Explicit construction of these observables and characterisation of their dynamical evolution can be achieved for simple totally constrained models. An immediate and tractable extension is to apply relational quantization to mini-superspace models. There, one has good reason to expect that the mutables prescription for observables combined with the extra freedom to take superpositions of energy eigenstates will allow for phenomena not permitted within the Dirac framework. This project will be taken up in future work. 

Before concluding we should again emphasise an important qualification regarding the limits of our proposal. Our approach is based upon the assumption that, when they exist, state symmetries can be unambiguously identified at the level of phase space. This is explicitly not the case for theories with refoliation symmetry such as General Relativity. Without further articulation, refoliation symmetries cannot be fit into either the $\gone$ or $\gtwo$ categories. The issue with refoliation symmetry does not, however, entirely bar the application of relational quantization to gravity. This is because, as noted in \cite{gryb:2011,gryb:2014}, the technique can be applicable to the Shape Dynamics formulation of gravity \cite{gryb:shape_dyn}. In this formulation, gravitation is no longer foliation invariant. The symmetries can be clearly distinguished as $\gone$ spatial diffeomorphism and Weyl symmetries, and $\gtwo$ reparamterization symmetry. Explicit details of this approach and further consideration of the issues of quantization and refoliation invariance will also be addressed in future work.

\section*{Funding}

K.T. would like to thank the Alexander von Humboldt Foundation and the Munich Center for Mathematical Philosophy (Ludwig-Maximilians-Universit\"{a}t M\"{u}nchen) for support. S.G. would like to acknowledge support from the Netherlands Organisation for Scientific Research (NWO) (Project No. 620.01.784) and Radboud University.

\section*{Acknowledgements}

We are appreciative to Oliver Pooley, Henrique Gomes and Lee Smolin for valuable discussion, and to Edward Anderson and Martin Bojowald for detailed critical comment on an earlier draft.

\clearpage

\bibliographystyle{utphys}
\bibliography{mach,Masterbib}

\end{document}